\documentclass{elsart}
\usepackage{mathrsfs}
\usepackage{amsfonts}

 \usepackage{epsfig}

\usepackage{amssymb}
\usepackage{amsmath}

\begin{document}

\begin{frontmatter}

\title{\large{\bf Evolution of Chinese airport network }}

\author{Jun Zhang$^{a}$, Xian-Bin Cao$^{a,b,*}$, Wen-Bo Du$^{a,b}$, Kai-Quan Cai $^{a}$}

\corauth[1]{Corresponding author. E-mail:
xbcao@ustc.edu.cn}
\address
{$^a$School of Electronic and Information Engineering, Beihang University, Beijing, 100083, P.R.China\\
$^b$School of Computer Science and Technology, University of
Science and Technology of China, Hefei, Anhui, 230026, P. R. China\\
}

\begin{abstract}
With the rapid development of economy and the accelerated
globalization process, the aviation industry plays more and more
critical role in today's world, in both developed and developing
countries. As the infrastructure of aviation industry, the airport
network is one of the most important indicators of economic
growth. In this paper, we investigate the evolution of Chinese
airport network (CAN) via complex network theory. It is found that
although the topology of CAN remains steady during the past
several years, there are many dynamic switchings inside the
network, which changes the relative relevance of airports and
airlines. Moreover, we  investigate the evolution of traffic
flow (passengers and cargoes) on CAN. It is found that the traffic
keeps growing in an exponential form and it has evident seasonal
fluctuations. We also found that cargo traffic and passenger traffic are positively
related but the correlations are quite different for different
kinds of cities.

\end{abstract}

\end{frontmatter}

\section{Introduction}

Ranging from biological systems to economic and social systems, many
real-world complex systems can be represented by networks, including
chemical-reaction networks, neuronal networks, food webs, telephone
network, the World Wide Web, railroad and airline routes, social
networks and scientific-collaboration networks
\cite{network1,network2,network3}. Obviously, the real networks are
neither regular lattices nor simple random networks. Since the
small-world network model \cite{WSmodel} and the scale-free network
model \cite{BAmodel} were brought forward at the end of the last
century, people find that many real complex networks are actually
associated with small-world property and a scale-free, power-law
degree distribution. In the past ten years, the theory of complex
networks has drawn continuous attention from different scientific
communities, such as network modelling \cite{model1,model2,model3},
synchronization \cite{synchronization1,synchronization2},
information traffic \cite{traffic1,traffic2,traffic3,traffic4},
epidemic spreading \cite{epidemic1,epidemic2}, cascading failures
\cite{cascade1,cascade2,cascade3,cascade4}, evolutionary games
\cite{game1,game2,game3,game4,game5} and social dynamics
\cite{social} etc.. One interesting and important research direction
is understanding the transportation infrastructures in the framework
of complex network theory
\cite{real1,real2,real3,real4,real5,real6,real7,real8}.

With the acceleration of globalization process, the aviation
industry plays a more and more critical role in the economy and
many scientists pay special attention to the airway transportation
infrastructure. Complex network theory is naturally a useful tool
since the airports can be denoted by vertex and the flights can be
denoted with edges. In the past few years, some interesting
researches have been reported to study the airport networks from
the view of network theory. For example, Amaral et al.
comprehensively investigated the worldwide airport network (WAN).
They found that WAN is a typical scale-free small-world network
and the most connected nodes in WAN are not necessarily the most
central nodes, which means critical locations might not coincide
with highly-connected hubs in the infrastructures. This
interesting phenomenon inspired them to propose a
geographical-political-constrained network model
\cite{Amaral1,Amaral2}. Vespignani et al. further investigated the
intensity of WAN's connections via the view of weighted networks
and they found the correlations between weighted quantities and
the topology. They proposed a weighted evolving network model to
expand our understanding of weighted features of real systems.
Besides, they also proposed a global epidemic model to study the
role of WAN in the prediction and predictability of global
epidemics \cite{Vespignani1,Vespignani2}. Besides, several
empirical works on Chinese Airport Network \cite{CAN1,CAN2,CAN3}
and Indian Airport Network \cite{IAN1} reveal that the scale of
national airport networks can exhibit different properties from
the global scale of WAN, i.e., the two-regime power-law degree
distribution and the disassortative mixing property.

As the aviation industry is an important indicator of economic
growth, it is necessary and more meaningful to investigate the
evolution of airport network. Recently, Gautreau et al. studied
the US airport network in the time period $1990 \sim 2000$. They
found that most statistical indicators are stationary and an
intense activity takes place at the microscopic level, with many
disappearing/appearing links between airports \cite{Gautreau1}.
Rocha studied the Brazilian airport network (BAN) in the time
period $1995 \sim 2006$. He also found the network structure is
dynamic with changes in the relevance of airports and airlines,
and the traffic on BAN is doubled during the period while the
topology of BAN shrinks \cite{Rocha1}. Inspired by their
interesting works, we investigate evolution of Chinese Airport
Network (CAN) from the year $1950$ to $2008$ ($1991$ to $2008$ for
detailed traffic information and $2002$ to $2009$ for detailed
topology information). It is found that the airway traffic volume
increases in an exponential form while the topology has no
significant change.

The paper is organized as follows. In the next section, the
description of CAN data is presented. The statistical analysis of
CAN topology is given in Section 3. In Section 4, we analyze
evolution of traffic flow on CAN. The paper is concluded by the
last section.

\section{Development of CAN with Chinese GDP}

\begin{figure}[htbp]
\centering {\psfig{file=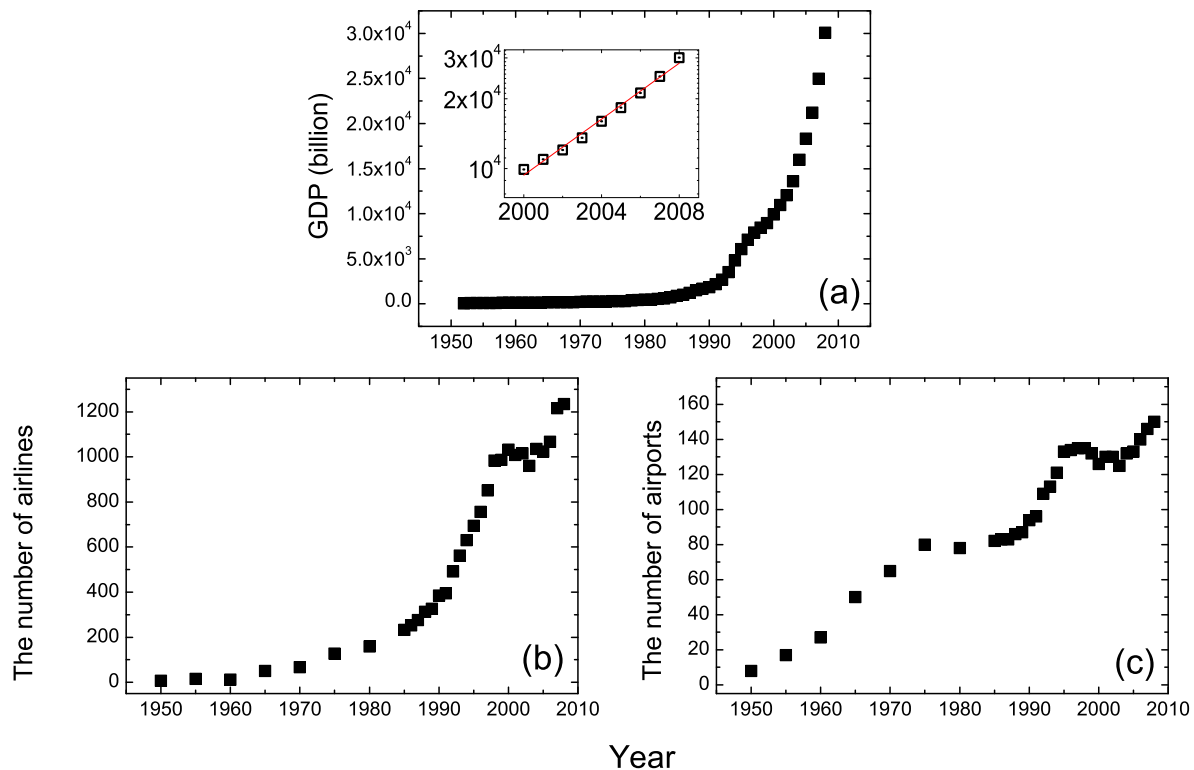,height=12cm}}
\caption{(Color online.) The development of Chinese Gross Domestic
Product (a), number of airlines (b) and number of airports (c)
from $1950$ to $2008$. The data are obtained from Ref.\cite{CAAC}}
\end{figure}

\begin{figure}[htbp]
\centering {\psfig{file=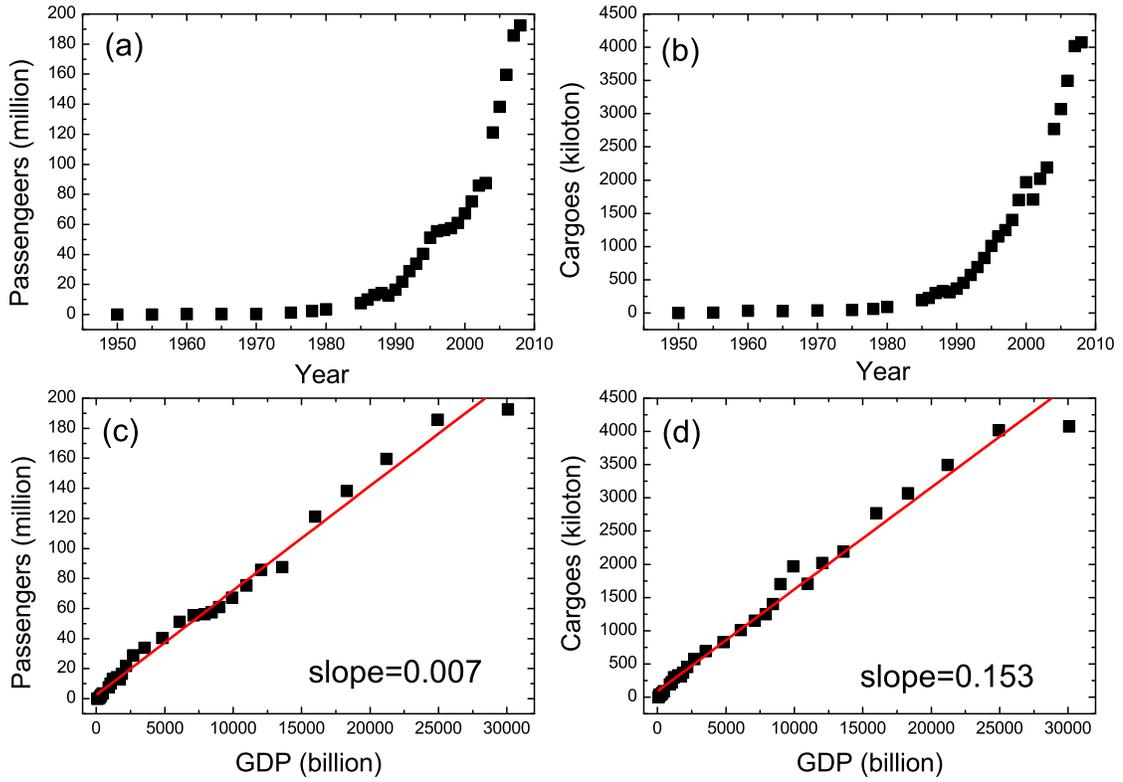,height=12cm}}
\caption{(Color online.) (a): The development of passengers,(b): The
development of  cargoes, (c): Relation of passengers over GDP, (d):
Relation of cargoes over GDP. The data are obtained from
Ref.\cite{CAAC} }
\end{figure}

Airport network is the backbone of aviation industry. It includes
airports and direct flights linking airport pairs. Since aviation
industry is closely related to economy development and China has
made a great economic miracle in the past decades, we firstly
investigate the development of Chinese economy, airports and
flights. Figure 1(a) shows the development of Chinese GDP from
$1950$ to $2008$. One can see that it has great increment in the
$58$ years. Especially, the historic Third Plenary Session of the
Eleventh Central Committee was held in $1978$, ushering in China's
new historical period of reform and opening up. Since then, Chinese
GDP increases faster and boosts in the beginning of $21st$ century
(GDP increases as an exponential form since year $2000$, see the
inset of Fig.1(a)). However, the development of airlines (Fig.1(b))
and airports (Fig.1(c)) is not in consistent with that of GDP. For
the development of airports (Fig.1(c)), one can see that the number
of airports grows in $1950 \sim 1975$, $1987 \sim 1995$  and $2005
\sim 2008$, but keeps constant in $1975 \sim 1987$ and $1995 \sim
2005$. The first increasing ($1950 \sim 1975$) mainly makes large
prefecture-level cities connected, and the second increasing ($1987
\sim 1995$) mainly makes medium prefecture-level cities connected.
The third increasing ($2005 \sim 2008$) is due to the rapid
development of Chinese economy and China plans to build more
airports by $2020$. From Fig.1(b), one can also see that the number
of airlines remains constant since $1995$ and rises again in year
$2007$ and $2008$. The steadiness is mainly due to efficiency
reason. Opening new airlines means more operating expenses and
commercial airline companies prefer to have a small number of hubs
where all airlines connect. They would not like to add uneconomical
airlines once a mature transportation network is constructed. Thus
the number of airlines does not increase continuously. In year
$2007$ and $2008$, as many new airports are put into service, many
new airlines are naturally launched.

Although the airline infrastructure (e.g., airports and airlines)
does not keep growing due to various constraints, the traffic on CAN
keeps growing with the GDP. As shown in Figure 2, the traffic
(passengers and cargoes) grows almost linearly with GDP. By
calculation, one can see that $1$ million RMB of GDP can support
about $7$ passengers and $153$ kg cargoes. Moreover, the Chinese
aviation industry is also shocked by the $2008$ global financial
crisis.  The top $3$ Chinese airline companies have reported their
operating information of $2008$ and most important indicators are
declining. This has been demonstrated by the annual report of Civil
Aviation Administration of China (CAAC) and  we can find in Fig.2
that the traffic of $2008$ is almost the same as that of $2007$.

\section{Topological properties of CAN}

The topology data of CAN are obtained from $14$ timetables
provided by Civil Aviation Administration of China (CAAC) from
$2002$ to $2009$ ($2$ timetables for years $2003\sim2008$, and $1$
timetable for the second half of $2002$ and the first half of
$2009$). It should be noted that:

\begin{itemize}
\item The timetable contains both domestic and international
airlines. As we only focus on the domestic information, the
international airlines are excluded.

\item Since Ref.\cite{CAAC} is a statistical yearbook edited by
CAAC, it contains not only the scheduled flights but also the
temporary flights, whereas the timetables only comprise the
scheduled flights. Thus the number of airlines in the timetable is
smaller than the data in Ref.\cite{CAAC} by about $150$.

\item Airports in one city are view as one airport. For instance, there are $3$
airports in Shanghai and Chengdu, and $2$ airports in Beijing.

\item The timetables are not perfectly in consistent with real
flights due to weather or emergencies.
\end{itemize}

\begin{figure}[htbp]
\centering {\psfig{file=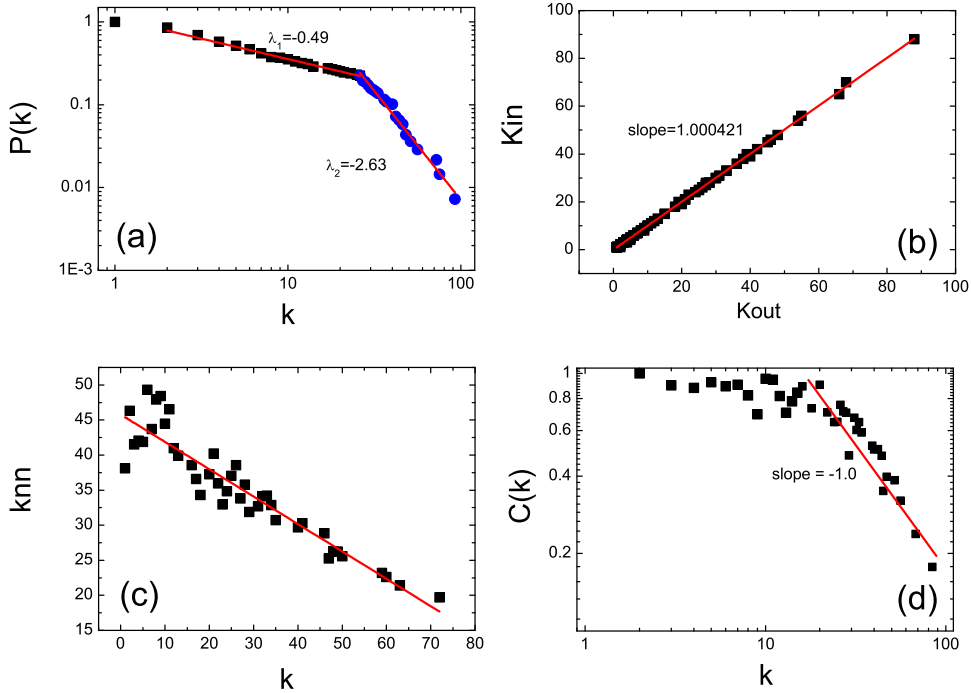,height=10cm}} \caption{(Color
online.) (a) Degree distribution of CAN;(b) Correlation between
$k_{in}$ and $k_{out}$ of CAN; (c) The degree-degree correlation of
CAN; (d) The clustering-degree correlation of CAN in the first half
year of $2009$. }
\end{figure}

Figure 3 shows some  basic topological characteristics of CAN in the
first half year of $2009$. Fig.3(a) shows the degree distribution
$P(k)$ of CAN, which follows a two-regime power-law distribution
with two different exponents ($\lambda_1 = -0.49$ and $\lambda_2 =
-2.63$ ). We also investigated the directed CAN and it is found that
$P(k_{in})$ and $P(k_{out})$ are almost the same as $P(k)$, where
$k_{in}$ is the ingoing degree and $k_{out}$ is the outgoing degree.
Fig.1(b) shows  the correlation between $k_{in}$ and $k_{out}$. One
can see that the in-out degree correlation is very strong: the slope
is $1.000421$. This means that one can fly from one airport to
another and return using the same airline. Another important
topological property is the degree-degree correlation. It is defined
as the mean degree of the neighbors ($k_{nn}$) of a given airport as
a function of the degree of the given airport. Fig.3(c) shows the
results of degree-degree correlation of undirected CAN and we can
find that the degrees of adjacent airports have significant linear
anti-correlation. Fig.3(d) exhibits the relationship of clustering
coefficient $C$ and degree $k$. As it shows, CAN has a power-law
decay of $C(k)$ as a function of degree $k$ ($C(k)\sim k^{-1}$),
which means that CAN is a hierarchical network and lower degree
nodes have larger clustering coefficient. All the results above are
well in accordance with the results reported by Liu et.al. and Li
et.al \cite{CAN1,CAN2,CAN3}.

In networks, a node participating more shortest paths is usually
more important. Thus the betweenness is proposed to quantify node's
importance in traffic \cite{Freeman1}. Figure 4 shows the relation
between degree and betweenness. One can see that betweenness
generally obeys an exponential function of degree but there exist
three nodes whose betweenness is obviously much larger: $Urumqi$,
$Xi'an$ and $Kunming$. The three nodes are all located in west
China: $Kunming$ is the central city of southwest, $Xi'an$ is the
central city of northwest and $Urumqi$ is the central city of far
northwest. The western population needs to be connected to the
political centers (e.g., Beijing) and economical centers (e.g.
Shanghai and Shenzhen) in the east. However, due to the long
distance from western China to eastern China (over $3,000$
kilometers), it is costly and unnecessary to make all western
airports directly link to the eastern airports. Thus some transit
airports are naturally formed as the bridge between east and west
China.

\begin{figure}[htbp]
\centering {\psfig{file=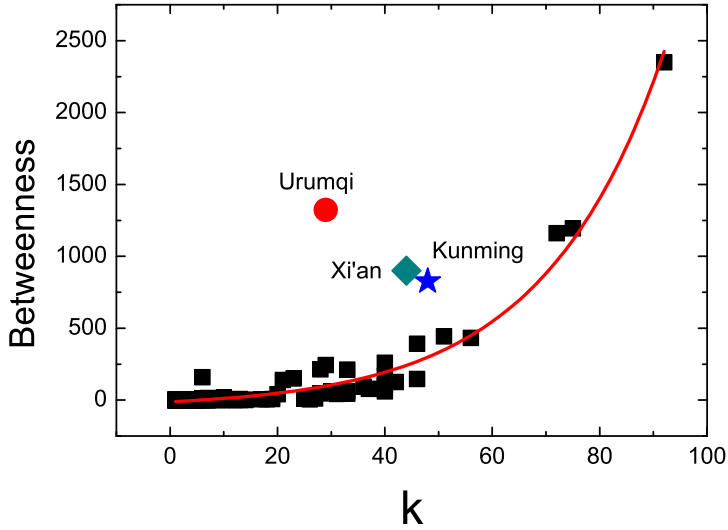,height=8cm}} \caption{(Color
online) The degree-betweenness correlation of CAN in the first
half year of $2009$. The fitting function is $y = A*e^{x/t_1}+y_0$
with $A = 42.0$, $t_1 = 22.6$ and $y_0 = -53.9$}
\end{figure}

Now we study evolution of topological properties of CAN. It can be
seen from Table 1 that topological properties of CAN do not
significantly change from $2002$ to $2009$. Similarly, topological
properties of Brazilian airport network also do not significantly
change during a long period of time \cite{Rocha1}. Next we make a
comparison between the two networks.

\begin{table}
\begin{center}
\caption{Evolution of topology parameters of CAN from year $2002$ to
$2009$. $\langle k \rangle$ is average degree, $k_{in}$ is ingoing
degree, $k_{out}$ is  outgoing degree, $d$ is average shortest path
length, $D$ is network diameter, $C$ is clustering coefficient, $R$
is a reciprocity parameter to measure the asymmetry of directed
networks and is defined as $R =\frac{\sum_{i\neq
j}^{N}(a_{ij}-\bar{a})(a_{ji}-\bar{a})}{\sum_{i\neq
j}^{N}(a_{ij}-\bar{a})^{2}}$ with $\bar{a} =\frac{\sum_{i\neq
j}^{N}{a_{ij}}}{N(N-1)}$ \cite{Rocha1}. Here $a_{ij}=1$ if there is
direct flight from airport $i$ to $j$, otherwise $a_{ij}=0$. }
\end{center}
\label{tab:data}
\begin{center}
\begin{tabular}{|c|c|c|c|c|c|c|c|c|c|c|c|c|c|} \hline
   Year &  $\langle k \rangle$ &  $\lambda_1$ & $\lambda_2$ & $k_{in}$ & $k_{out}$ & $R$ & $C$ & $d$ & $D$  \\ \hline
    2002(2) &  13.90 & -0.42 & -2.66 & 13.78 & 13.78 &0.990 & 0.75 & 2.21 & 5  \\
    2003(1) & 12.85 & -0.44 & -2.79 & 12.71 & 12.71  &0.988 & 0.70 & 2.24 & 5  \\
    2003(2) &  11.81 & -0.41 & -2.63 & 11.69 & 11.69 &0.989 & 0.71 & 2.26 & 5  \\
    2004(1)& 12.78 & -0.43 & -2.58 & 12.68 & 12.68  &0.991 & 0.75 & 2.22 & 4  \\
    2004(2) & 11.70 & -0.45 & -2.53 & 11.61 & 11.61 &0.991 & 0.77 & 2.23 & 4  \\
    2005(1) &  11.55 & -0.45 & -2.67 & 11.24 & 11.24 &0.970 & 0.79 & 2.27 & 4  \\
    2005(2)&  12.03 & -0.45 & -2.52 & 11.90 & 11.90 &0.988 & 0.79 & 2.25 & 4  \\
    2006(1) &  11.71 & -0.47 & -2.77 & 11.66 & 11.66 & 0.995 & 0.77 & 2.28 & 4  \\
    2006(2) &  12.55 & -0.46 & -2.81 & 11.94 & 11.94 &0.944 & 0.81 & 2.22 & 4  \\
    2007(1) &  12.33 & -0.45 & -2.52 & 12.23 & 12.23 &0.991 & 0.79 & 2.28 & 4 \\
    2007(2) &  12.85 & -0.47 & -2.96 & 12.88 & 12.88 &0.994 & 0.79 & 2.25 & 4  \\
    2008(1) &  13.22 & -0.47 & -2.64 & 13.38 & 11.37 &0.990 & 0.78 & 2.23 & 4  \\
    2008(2) &  12.06 & -0.46 & -2.70 & 11.96 & 11.96 &0.991 & 0.76 & 2.29 & 4  \\
    2009(1) &  13.07 & -0.49 & -2.63 & 12.97 & 12.97 &0.991 & 0.79 & 2.27 & 4  \\ \hline
\end{tabular}
\end{center}
\end{table}

\begin{figure}[htbp]
\centering {\psfig{file=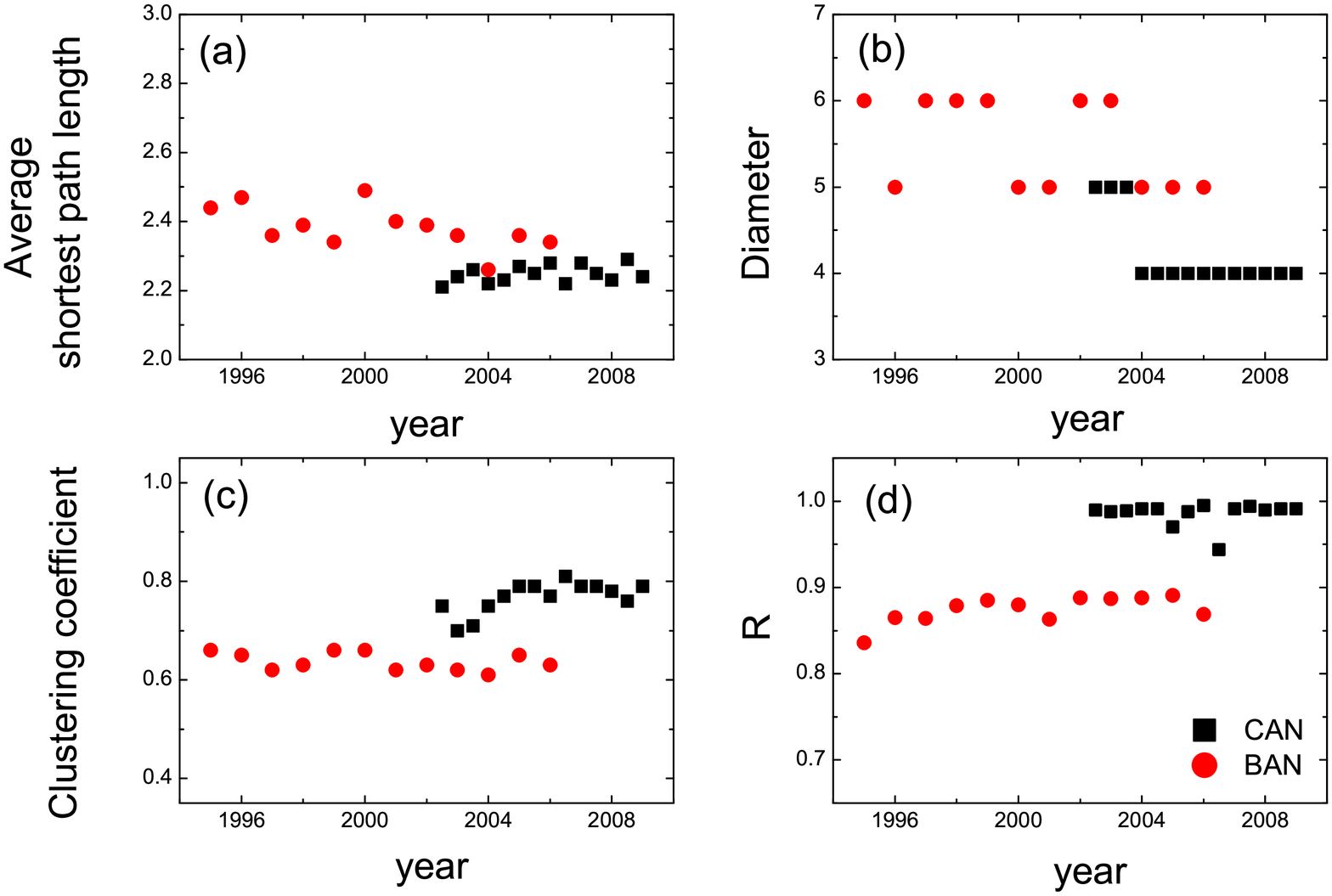,height=10cm}}
\caption{(Color online.) (a) The average shortest path length $d$,
(b) The diameter $D$, (c) The clustering coefficient $C$, (d) The
reciprocity parameter $R$ of CAN and BAN. The data of BAN is
reproduced from Ref.\cite{Rocha1}.}
\end{figure}

Fig.5(a) compares average shortest path length $d$ of CAN and BAN.
One can see that $d$ of CAN is around $2.25$ and is slightly smaller
than that of BAN. Fig.5(b) shows the diameter $D$, which is also
slightly smaller in CAN. This means that CAN is more convenient for
passengers. Table $2$ gives detailed results of shortest paths of
CAN in the first half year of $2009$. About $10\%$ paths are direct
connections and over $98\%$ paths consist of no more than $2$
flights. Fig.5(c) shows that average clustering coefficient $C$ of
CAN is apparently larger than that of BAN and Fig.5(d) shows that
CAN is more reciprocal than BAN.

\begin{table}
\begin{center}
\caption{Distribution of shortest paths in the first half year of
$2009$ }
\end{center}
\label{tab:data}
\begin{center}
\begin{tabular}{|c|c|c|c|} \hline
    Shortest Path &  Number of Paths &  Percentage of Paths & Number of Flights to be changed   \\ \hline
    1 & 902 & 9.54 & 0  \\ \hline
    2 & 5561 & 58.83 & 1   \\ \hline
    3 & 2853 & 30.18 & 2  \\ \hline
    4 & 137 & 1.45 & 3    \\ \hline
\end{tabular}
\end{center}
\end{table}

\begin{figure}[htbp]
\centering {\psfig{file=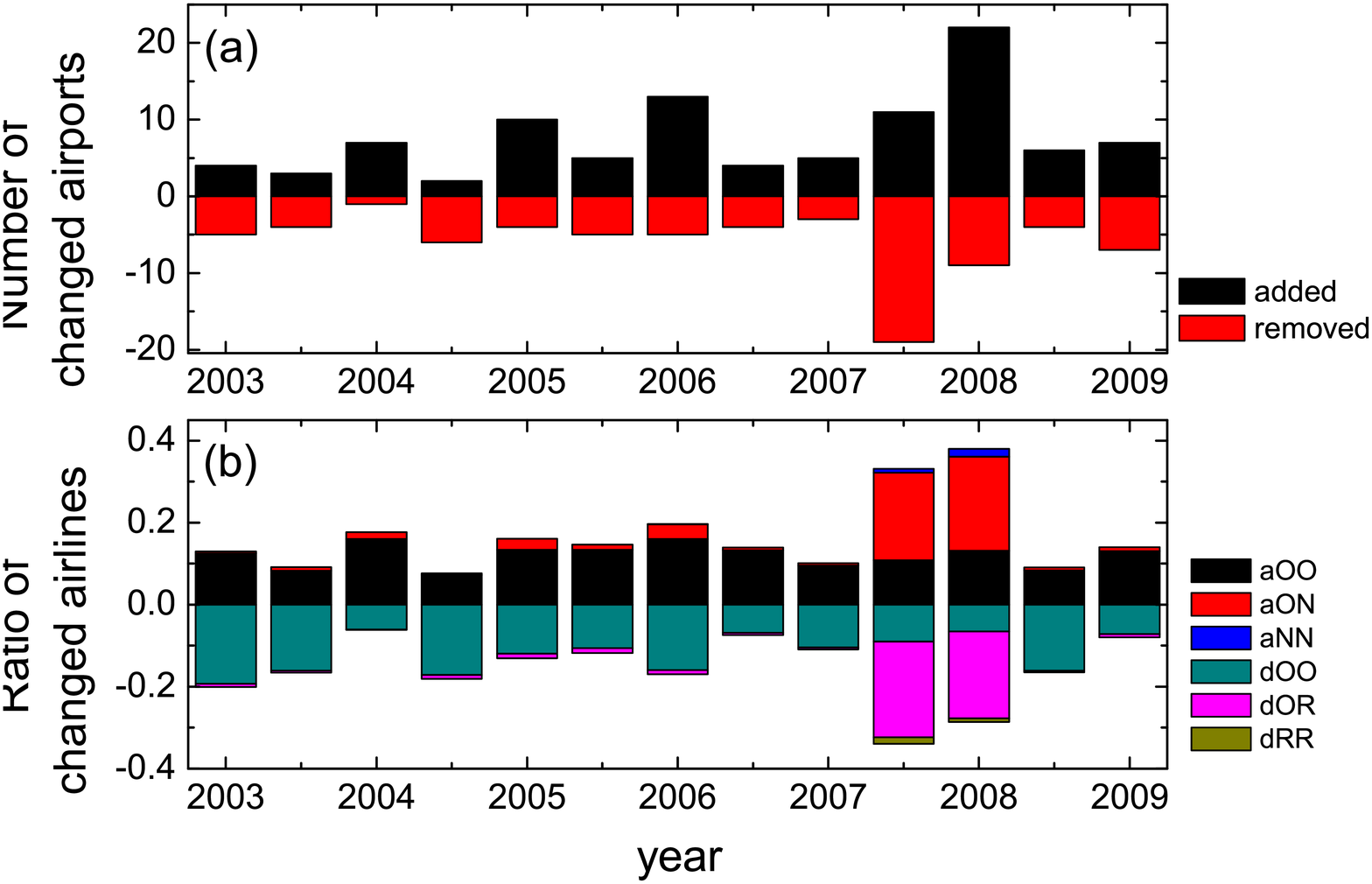,height=8cm}}
\caption{(Color online.) (a)  The fluctuation of airports: $added$
indicates the new airports and $removed$ indicates the removed
airports; (b) the fluctuation of airlines: $aOO$ indicates the
added airlines between old airports, $aON$ indicates the added
airlines between old and new airports,  $aNN$ indicates the added
airlines between  new airports,  $dOO$ indicates the deleted
airlines between old airports,  $dOR$ indicates the deleted
airlines between old and removed airports and $dRR$ indicates the
deleted airlines between removed airports.}
\end{figure}

From discussions above, we know that CAN is an asymmetric
small-world network with a two-regime power-law degree distribution,
a high clustering coefficient, a short average path length, a
negative degree-degree correlation, a negative clustering-degree
correlation and an exponential betweenness-degree correlation.
Although the topology characteristics of CAN is quite steady from
year $2002$ to $2009$, a dynamic switching process underlies the
evolution of CAN. Figure 6 shows the measured fluctuation of CAN
from year $2002$ to $2009$. Fig.6(a) shows the fluctuation of
airports and we can see that the fluctuation (including the added
airports and removed airports) is usually between $5$ and $15$. But
for the second half year of $2007$ and the first half year of
$2008$, the fluctuation is evidently more vigorous. Fig.6(b) shows
that the percentage of changed airlines is usually smaller than
$20\%$ and the majority of changes is mainly induced by $aOO$ and
$dOO$. But for the second half year of $2007$ and the first year of
$2008$, when many airports were added and removed, $aON$ and $dOR$
becomes the majority of changes.

\section{The traffic of CAN}

\begin{figure}[htbp]
\centering {\psfig{file=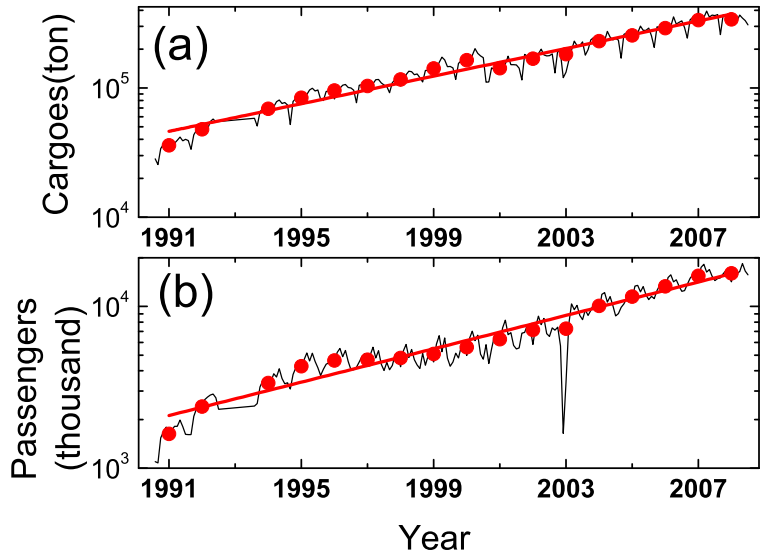,height=8cm}}
\caption{(Color online) Evolution of total traffic on CAN: (a)
cargoes; (b) passengers.}
\end{figure}

\begin{figure}[htbp]
\centering {\psfig{file=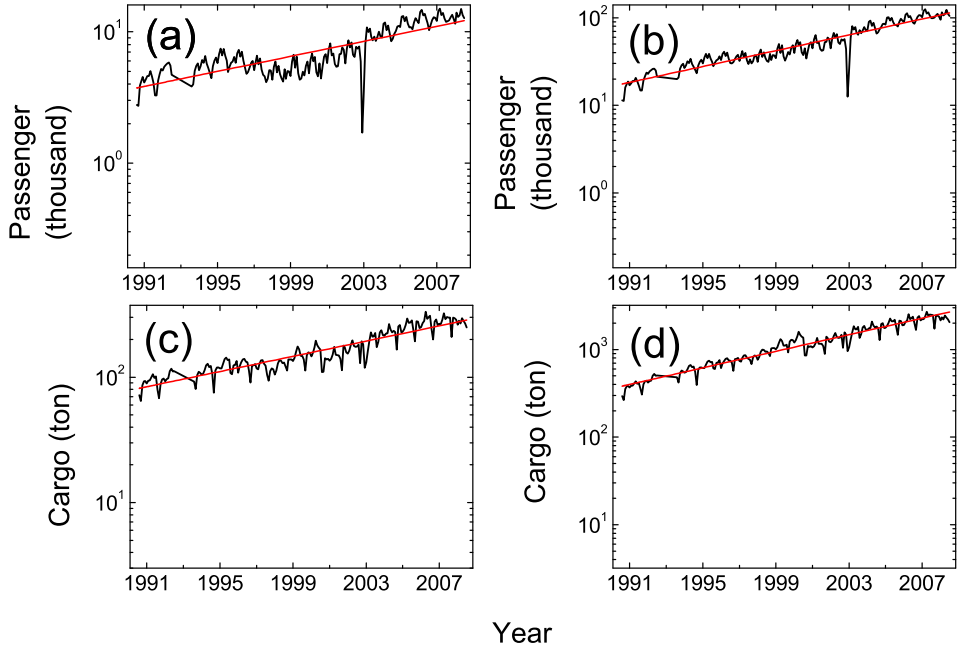,height=10cm}}
\caption{(Color online)  Evolution of total traffic on CAN: (a)
passengers per-link; (b) passengers per-node; (c) cargoes per-link;
(d) cargoes per-node.}
\end{figure}

\begin{figure}[htbp]
\centering {\psfig{file=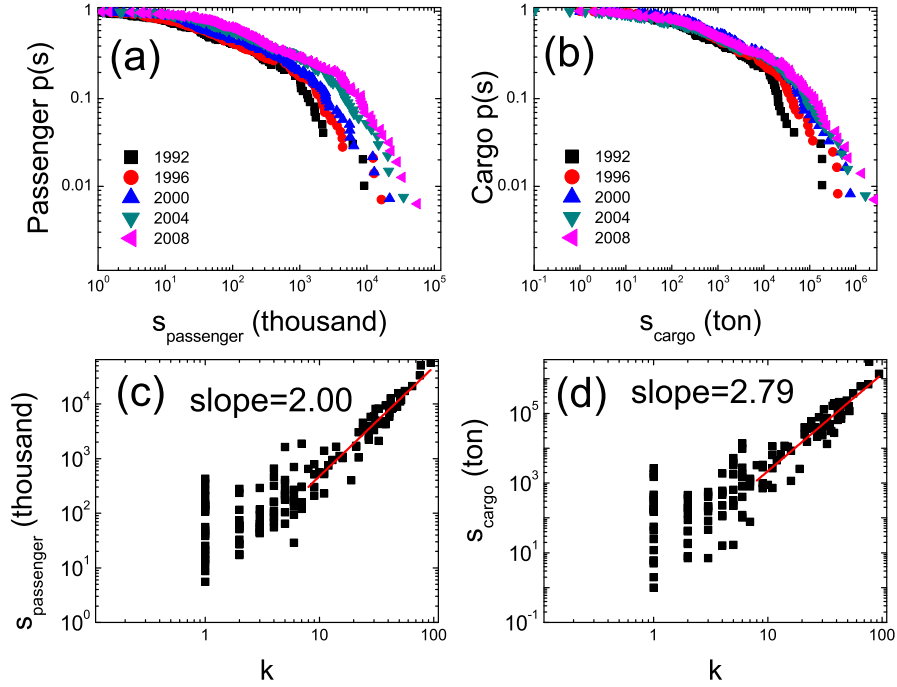,height=10cm}} \caption{(Color
online)(a)$s_{passenger}$: the passenger throughput of each airport,
(b)$s_{cargo}$: the cargo throughput of each airport of years
$1992$, $1996$,$2000$,$2004$ and $2008$;  (c) the correlation
between $k$ and $s_{passenger}$, (d) the correlation between $k$ and
$s_{cargo}$ in year $2008$.  }
\end{figure}

\begin{figure}[htbp]
\centering {\psfig{file=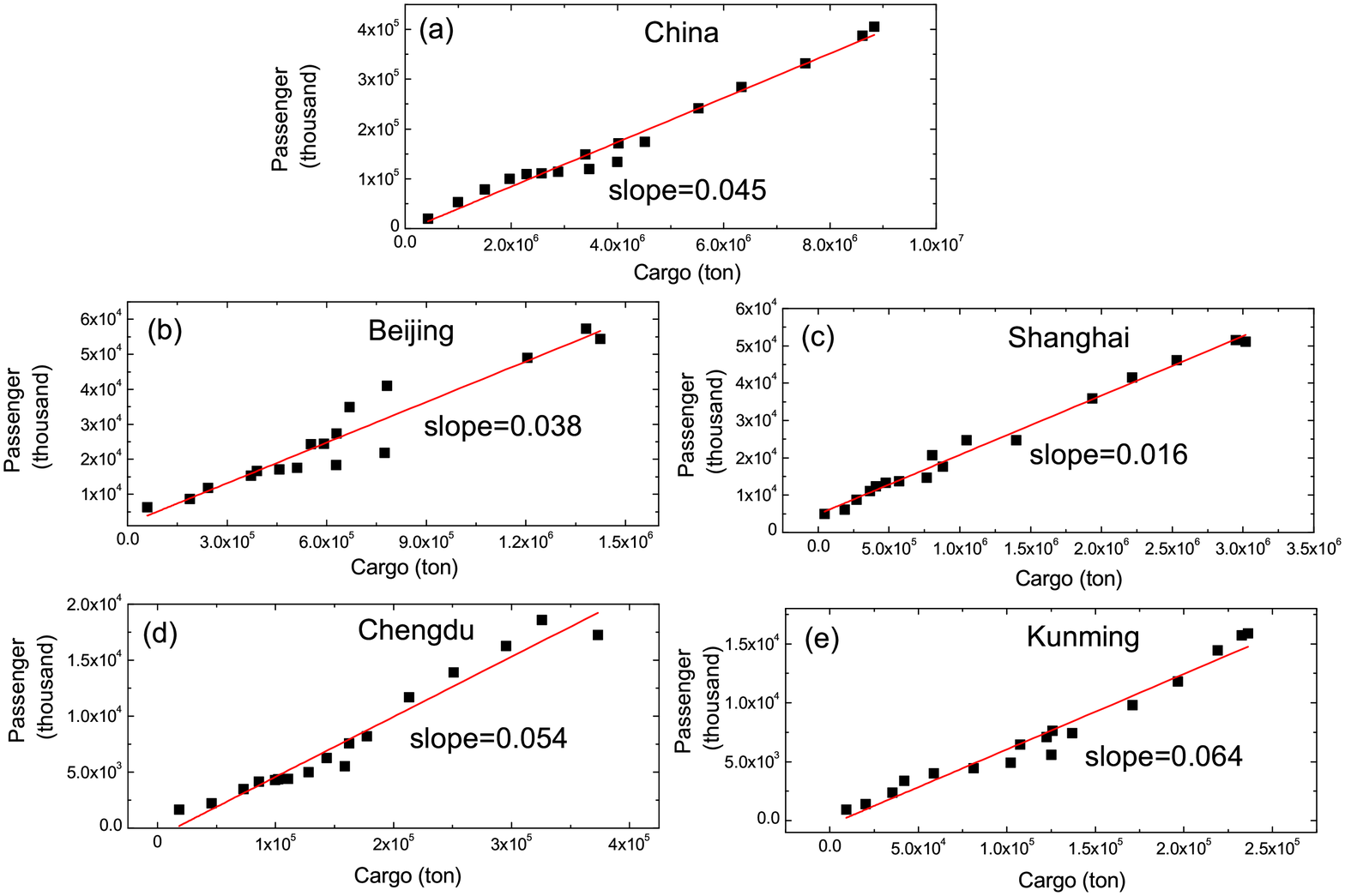,height=12cm}}
\caption{(Color online)  The correlations of cargo traffic and
passenger traffic from year $1991$ to $2008$ for: (a) the whole
country, (b) Beijing, (c) Shanghai, (d) Chengdu and (e) Kunming.}
\end{figure}

This section investigates evolution of traffic on CAN.

As shown in Figure 7,  the traffic (including cargoes and
passengers) has evident seasonal fluctuations as in the United
States. If the seasonal fluctuations are averaged out, one finds
that the traffic of CAN increases exponentially, much faster than
that of the United States (as shown in Ref.\cite{Gautreau1}, the
passenger data of the U.S. can be linearly fitted). We can also
observe similar growth (Figure 8) of the average traffic per-link
and per-node. It is found that the average traffic of CAN has
increased about $200\%$ during the $17$ years while the average
passenger traffic of the U.S. has only increased about $20\%\sim
35\% $ during the $10$ years ($1990$ to $2000$). It is worth noting
that there exists a sudden drop of passenger traffic in year $2003$
(see Fig.7(b), Fig.8(a) and Fig.8(b)). This is mainly induced by the
Severe Acute Respiratory Syndrome (SARS). However, the cargo traffic
was not knocked by SARS (see Fig.7(a), Fig.8(c) and Fig.8(d)).

Figure 9 displays the cumulative distribution of nodes' strength
$s$, namely the throughput of each airport including passengers
 ($s_{passenger}$, see Fig.9(a)) and  cargoes ($s_{cargo}$, see
Fig.9(b)). The distributions are quite broad: $5$ orders of
magnitude for passengers and $7$ for cargoes. The correlations of
$k$ and $s$ are also presented. Fig.9(c) shows the dependence  of
$s_{passenger}$ on $k$, and Fig.9(d) shows the dependence  of
$s_{cargo}$  on $k$ in year $2008$. One can find that there exists
a clear non-linear behavior denoting a strong correlation between
strength and topology: $s_{passenger} \propto k^{2.00}$ and
$s_{cargo} \propto k^{2.79}$. We also examined the data from year
$2002$ to $2007$ and the results are similar.

Figure 10 shows the correlations of cargo traffic and passenger
traffic from year $2001$ to $2008$. One can find a strong linear
correlation between  cargo traffic and passenger traffic for both
the total traffic of CAN and the traffic of a single airport/city.
However, the ratios of cargo traffic and passenger traffic are quite
different. As shown in Fig.10(a), the slope is $0.045$ for the total
traffic of CAN. For municipalities Beijing (Fig.10(b)) and Shanghai
(Fig.10 (c)), the slopes are obviously smaller. Because Beijing and
Shanghai are the most important central cities of politics and
economy and culture of China, they are aggregating centers and
distributing centers for over $51\%$ of Chinese goods flow (only
$27\%$ of Chinese passenger flow). For tourism cities Chengdu
(Fig.10(d)) and Kunming (Fig.10(e)), the  slopes are obviously
larger, indicating that the passenger traffic is more active in
these two cities.

\section{Conclusion}

In summery, we investigate the evolution of Chinese airport network
(CAN), including the topology, the traffic and the interplay between
them. We find that, though the main topological indicators are quite
stationary, there exists a dynamic switching process inside the
network (airports added and removed, airlines added and removed).
Moreover, the traffic flow (including passengers and cargoes) on CAN
is studied. The traffic grows at an exponential rate with seasonal
fluctuations, and the traffic throughput of an airport has a
nonlinear correlation with its degree. Moreover, our comparative
studies show that cargo traffic and passenger traffic are positively
related, but with different ratioes for different kinds of
cities. Our work provides insights in understanding the evolution of
national airport network.

\section*{Acknowledgement}
We thank Gang Yan, Rui Jiang and Mao-Bin Hu for their useful
discussions.  This work is supported by the National Science
Foundation for Distinguished Young Scholars of China (No.60625102)
and the Foundation for Innovative Research Groups of the National
Natural Science Foundation of China (No.60921001).

\end{document}